\documentclass[aps,prx,reprint,preprintnumbers,superscriptaddress,nofootinbib,longbibliography,floatfix]{revtex4-2}
\pdfoutput=1
\pdfoutput=1
\usepackage{rotating}
\usepackage{array}
\usepackage{amsmath}
\usepackage[normalem]{ulem}
\usepackage{slashed}
\usepackage{booktabs}
\usepackage[pdftex,table]{xcolor}
\usepackage{xfrac}
\usepackage{mathtools}
\usepackage{empheq}
\usepackage{multirow}
\usepackage{amssymb}
\usepackage{url}
\usepackage{comment}
\usepackage{physics}
\usepackage{color,soul}
\usepackage{csquotes}
\usepackage{lineno}
\usepackage[caption=false]{subfig}
\usepackage[normalem]{ulem}
\graphicspath{{Figures/}}
\usepackage{hyperref}

\hypersetup{
  colorlinks=true,
  citecolor=blue,
  linkcolor=blue,
  urlcolor=blue
}

\begin{document}

\title{The Optimal use of Segmentation for Sampling Calorimeters}

\author{Fernando Torales Acosta}
\email{ftoralesacosta@lbl.gov}
\affiliation{Physics Division, Lawrence Berkeley National Laboratory, Berkeley, CA 94720, USA}
\author{Bishnu Karki}
\email{bishnu.karki@ucr.edu}
\affiliation{Department of Physics and Astronomy, University of California, Riverside, CA 92521, USA}

\author{Piyush Karande}
%\email{karande1@llnl.gov}
\affiliation{Computational Engineering Division, Lawrence Livermore National Laboratory, Livermore CA 94550}

\author{Aaron Angerami}
%\email{angerami1@llnl.gov}
\affiliation{Nuclear and Chemical Science Division, Lawrence Livermore National Laboratory, Livermore, CA 94550}

\author{Miguel Arratia}
%\email{miguel.arratia@ucr.edu}
\affiliation{Department of Physics and Astronomy, University of California, Riverside, CA 92521, USA}
\affiliation{Thomas Jefferson National Accelerator Facility, Newport News, Virginia 23606, USA}
\author{Kenneth Barish}
%\email{kenneth.barish@ucr.edu}
\affiliation{Department of Physics and Astronomy, University of California, Riverside, CA 92521, USA}
%\affiliation{Brookhaven National Laboratory, Upton, New York, 11973, USA}

\author{Ryan Milton}
%\email{rmilt003@ucr.edu}
\affiliation{Department of Physics and Astronomy, University of California, Riverside, CA 92521, USA}
\author{Sebasti\'an Mor\'an}
%\email{miguel.arratia@ucr.edu}
\affiliation{Department of Physics and Astronomy, University of California, Riverside, CA 92521, USA}

\author{Benjamin Nachman}
%\email{bpnachman@lbl.gov}
\affiliation{Physics Division, Lawrence Berkeley National Laboratory, Berkeley, CA 94720, USA}
\affiliation{Berkeley Institute for Data Science, University of California, Berkeley, CA 94720, USA}

\author{Anshuman Sinha}
%\email{karande1@llnl.gov}
\affiliation{Computational Engineering Division, Lawrence Livermore National Laboratory, Livermore CA 94550}

\begin{abstract}
    One of the key design choices of any sampling calorimeter is how fine to make the longitudinal and transverse segmentation.  To inform this choice, we study the impact of calorimeter segmentation on energy reconstruction.  To ensure that the trends are due entirely to hardware and not to a sub-optimal use of segmentation, we deploy deep neural networks to perform the reconstruction.  These networks make use of all available information by representing the calorimeter as a point cloud.  To demonstrate our approach, we simulate a detector similar to the forward calorimeter system intended for use in the ePIC detector, which will operate at the upcoming Electron Ion Collider.  We find that for the energy estimation of isolated charged pion showers, relatively fine longitudinal segmentation is key to achieving an energy resolution that is better than 10\% across the full phase space.  These results provide a valuable benchmark for ongoing EIC detector optimizations and may also inform future studies involving high-granularity calorimeters in other experiments at various facilities.
\end{abstract}
\maketitle

%\tableofcontents

\section{Introduction}
\label{sec:intro}
Calorimeters are a critical component of many particle and nuclear physics detectors.  The reconstruction of the energy flow in such detectors requires custom algorithms because of the diversity of possible geometry and readout configurations.  For running experiments, these algorithms are optimized over years by a number of researchers.  Classically, this is not possible during the design phase and so significantly simplified approaches are used during this stage.  If the hardware is designed with suboptimal software, then the resulting instrument will be suboptimal.  This question is particularly acute for the calorimeter(s) of the Electron Ion Collider (EIC), as they are being designed now and will need to be finely segmented to deliver the science goals of the EIC~\cite{Accardi:2012qut,AbdulKhalek:2021gbh}.

We propose to employ machine learning (ML) to address this key challenge with traditional detector optimization.  Deep neural networks can readily process the lowest-level information within a cluster of cells produced from an electromagnetic or hadronic shower.  Previous studies on calorimeter energy estimation and particle identification for collider experiments have shown that utilizing this information outperforms existing approaches and provides an excellent approximation to the optimal reconstruction~\cite{deOliveira:2018hva,CERN-LHCC-2017-023,ATL-PHYS-PUB-2020-018,Neubuser:2021uui,Akchurin:2021afn,Akchurin:2021ahx,ATL-PHYS-PUB-2022-040,Qasim:2022rww,Kieseler:2021jxc,Neubuser:2021uui}.  Most importantly, machine learning approaches are constructed automatically, enabling hardware design to be guided by approximately optimal software utilization.  In particular, the development of software-based compensation for sampling calorimeters can be streamlined.  The goal is to extract the benefits of classical, complex algorithms like those studied with ATLAS~\cite{ATLAS:2016krp} and CALICE~\cite{CALICE:2012eac} with a training latency that allows for the comparison of many detector configurations in a reasonable amount of time.

While most of the literature is built on a fixed instrument, a couple of studies have started to explore the interplay between calorimeter design and machine learning-based energy estimation.  The convolutional neural network (CNN) in Ref.~\cite{Akchurin:2021afn} was deployed to different configurations of the cell energy calibration, noise, and bias.  A similar CNN in Ref.~\cite{Neubuser:2021uui} was used to systematically examine the relationship between bias/resolution and segmentation in a homogeneous hadronic calorimeter.  

In this paper, we study the optimal use of segmentation for sampling calorimeters by employing a variety of neural network architectures.  Previous studies have shown that point-cloud methods are more powerful than image-based approaches~\cite{ATL-PHYS-PUB-2022-040}.  While the calorimeters we study have the same transverse segmentation in each longitudinal layer (unlike the ATLAS calorimeter of Ref.~\cite{ATL-PHYS-PUB-2022-040}), we still expect point cloud methods to outperform CNNs given the sparsity of calorimeter images.  As such, we focus on Graph Neural Networks (GNNs) (see e.g. Ref.~\cite{Shlomi:2020gdn}) with and without (also called DeepSets~\cite{deepsets}) edges.  The calorimeters we study resemble the CALICE AHCAL~\cite{CALICE:2022uwn}, with a particular focus on configuration similar to the ePIC detector intended for use at the EIC~\cite{Bock:2022lwp,hcalInsert}.  Our objective is to offer insights into the ongoing design and optimization of the ePIC detector, considering the expected jet energies at the EIC.  In particular, the kinematic range for single particle energies is about tens of GeV at pseudo-rapidity $\eta=1$ to approximately 250 GeV at $\eta=3.5$~\cite{AbdulKhalek:2021gbh}.

This paper is organized as follows.  The ML approaches used for energy regression are described in Sec.~\ref{sec:ML}.  Our simulated detector is detailed in Sec.~\ref{sec:sim}.  Numerical results are presented in Sec.~\ref{sec:results}, and the paper ends with conclusions and outlook in Sec.~\ref{sec:conclusions}.

\section{Regression Models}
\label{sec:ML}
Our studies build upon the approach to hadronic energy calibration developed by the ATLAS Collaboration, which are described in Refs.~\cite{ATL-PHYS-PUB-2022-040,ATL-PHYS-PUB-2020-018}, and summarized in the following sections. 
\subsection{DeepSets}
\label{sec:DeepSets}
The DeepSets theorem stipulates that any observable that is symmetric with respect to the ordering of particles can be approximated arbitrarily well with a parameterization of permutation invariant functions of variable length inputs \cite{deepsets}.

This theorem can be applied in the context of calorimeter energy calibration, where observables are viewed as functions of clusters composed of calorimeter cells. A cluster is comprised of $N$ cells, or a set of \textit{nodes}, $V$, where $v_i$ contains the relevant features of cell $i$. The cell level features considered in this work are the logarithm of the cell energy, the $X$, $Y$, and $Z$ positions of the cell, and the calorimeter index $c$ described in Section \ref{sec:dataset}. For studies investigating the effect of longitudinal segmentation, only the logarithm of the sum of cell energies per layer in the HCAL is considered. Additionally, the total energy sum deposited in the the calorimeter is included as an input in the form of a \textit{global node}, $U$, and its feature $u$ which is relevant to the entire input set representing the cluster.

The general parameterization for a permutation-invariant observable is given by the following equations using the variables described above. The first step is the mapping of the node or cell level features to the latent representation $v_i'$ using the function $f_\mathrm{node}$:

\begin{equation}
        v_i' = f_\mathrm{node}(u, v_i).
        \label{eq:deepsets_nodes}
    \end{equation}

The second step is the aggregation of information from all nodes using a permutation-invariant function followed by mapping to another latent representation to describe the observable cluster using the function $f_\mathrm{global}$:

\begin{equation}
    u' = f_\mathrm{global}\left(\sum_{i\in\mathcal{N}}v'_i\right).
    \label{eq:deepsets_global_nodes}
\end{equation}

Fully connected (`dense') neural networks also known as Multi-Layer Perceptrons (MLPs) are used to approximate $f_\mathrm{node}$ and $f_\mathrm{global}$. Both of these networks use four dense layers with 64 nodes each, thus resulting in a latent space representation of size 64. Each dense layer uses the Rectified Linear Unit (ReLU) activation function~\cite{relu} and He-normal initialization \cite{HeUniform}. %The four dense layers together are labeled as Multi-Layer Perceptrons (MLPs). 

\begin{figure*}[tbh]
    \begin{center}
        \includegraphics[width=0.85\textwidth]{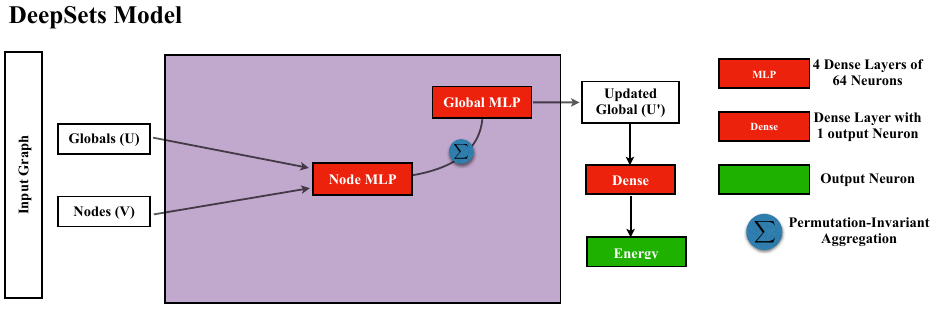}
    \end{center}
    \caption{A schematic of the DeepSets model.}
    \label{fig:DeepSets_Block}
\end{figure*}

Figure \ref{fig:DeepSets_Block} shows a basic schematic of the DeepSets model used in this work. It outlines the distinct MLPs for both nodes and global nodes used in the DeepSets model, the aggregation function, as well as the final output, Energy.

Each node is passed through the \textit{Node MLP} after the global node features are concatenated with it. The Node MLP approximates $f_\mathrm{node}$ as shown in Eq. \ref{eq:deepsets_nodes}. The output of the Node MLP is aggregated across all nodes and used as an input to the \textit{Global MLP} which approximates the $f_\mathrm{global}$ and is shown in Eq. \ref{eq:deepsets_global_nodes}. The aggregation of nodes and the global features define the DeepSets block. Finally, the output of the \textit{Global MLP} is passed through a dense layer to predict the output Energy. 

The model is trained with a batch size of 256 calorimeter showers for 100 epochs, using the Adam optimizer~\cite{adam}. The learning rate was initialized to $1e^{-3}$ and was halved every 5 epochs to a minimum of $1e^{-6}$. Mean Absolute Error (MAE) loss was used\footnote{Note that direct regression is prior dependent in the sense of Ref.~\cite{Gambhir:2022dut}.  We expect this effect to be numerically small and will explore it in future studies.}.

For the first set of results, DeepSets models are used to investigate the impact of progressively higher-fidelity cell information in Section \ref{sec:results}, labeled 1D, 2D, and 4D DeepSets models. For the 1D DeepSets model, only the energy of each cell is used to train the model. In the 2D DeepSets model, the cell energy and $Z$ position information are used. The 4D DeepSets model uses Energy, $Z$, $X$, and $Y$ position information.

    \subsection{Graph Neural Networks}
    \label{sec:GNN}
    Another model used in this work and compared to the DeepSets results is the GNN model. GNNs are very similar to DeepSets; they can take as input unordered set of cells to construct permutation invariant observables. They similarly use the same concepts of \textit{nodes} and \textit{global nodes} described in \ref{sec:DeepSets}. While both GNNs and DeepSets are well suited for the task of energy reconstruction, GNNs take as input additional information in the form of \textit{edges}. For this work, edges represent the set of nearest neighbors of a cell, in addition to the cell's node features, the standard cell information used in the DeepSets models (cell $E, X, Y,$ and $Z$). Each node, represented as a cell and its relevant attributes, is given ten edges, represented as the distance between the cell and the ten nearest cells. In this way, local information about the shower geometry is directly encoded as input to the GNN model.

    A typical form of a GNN takes a graph-structure input $G=(U, V, E)$, where once again $U$ is the \textit{global node}, inherently connected to all the \textit{nodes} in the set $V$, which are all connected to each other with the set of \textit{edges}, $E$. The GNN learns a hidden representation of the graph that is updated via a three-step message-passing process outlined in Ref.\cite{GNN_training}. This message-passing process involves functions of the global node features $u$, the node features $v_i$, and the edge features $e_{ij}$. The updated edge features $e_{ij}'$ are defined as:
    
    \begin{equation}
            e_{ij}' = f_\mathrm{edge}\left(u, v_i, v_j, e_{ij}\right).
            \label{eq:msg_gnn_edge}
    \end{equation}
    
    This first step in the message passing process, \textit{updates} all the edges to $e'_{ij}$, using the global node feature $u$, current edge feature $e_{ij}$, and the features from the two nodes connected by them, $v_i$ and $v_j$. The updated node features $v_i'$ are defined as:
    
    \begin{equation}
        v_i' = f_\mathrm{node}\left(u, v_i, \sum_{j\in\mathcal{N}_i}e_{ji}'\right).
        \label{eq:msg_gnn_nodes}
    \end{equation}
    
     This second step \textit{aggregates} the information from the target node's neighbors, $\mathcal{N}_i$ to the target node, using a permutation-invariant function ($\sum$), similar to those used in DeepSets. Each node is then \textit{updated} to $v_i'$ by using the global node feature $u$, current node features $v_i$ and the ``messages" aggregated from its neighbors in the updated edge features $e_{ij}'$ to form an embedded representation.

The last step is to aggregate the hidden node embeddings in order to quantify the entire graph structure for the purpose of energy regression. This aggregation of node embeddings is performed by updating the global node features $u$ that encode graph-level attributes:

\begin{equation}
    u' = f_\mathrm{global}\left(u, \sum_{i\in\mathcal{N}}v'_i\right).
    \label{eq:gnn_global_nodes}
\end{equation}

\begin{figure*}[tbh]
        \begin{center}
            \includegraphics[width=0.85\textwidth]{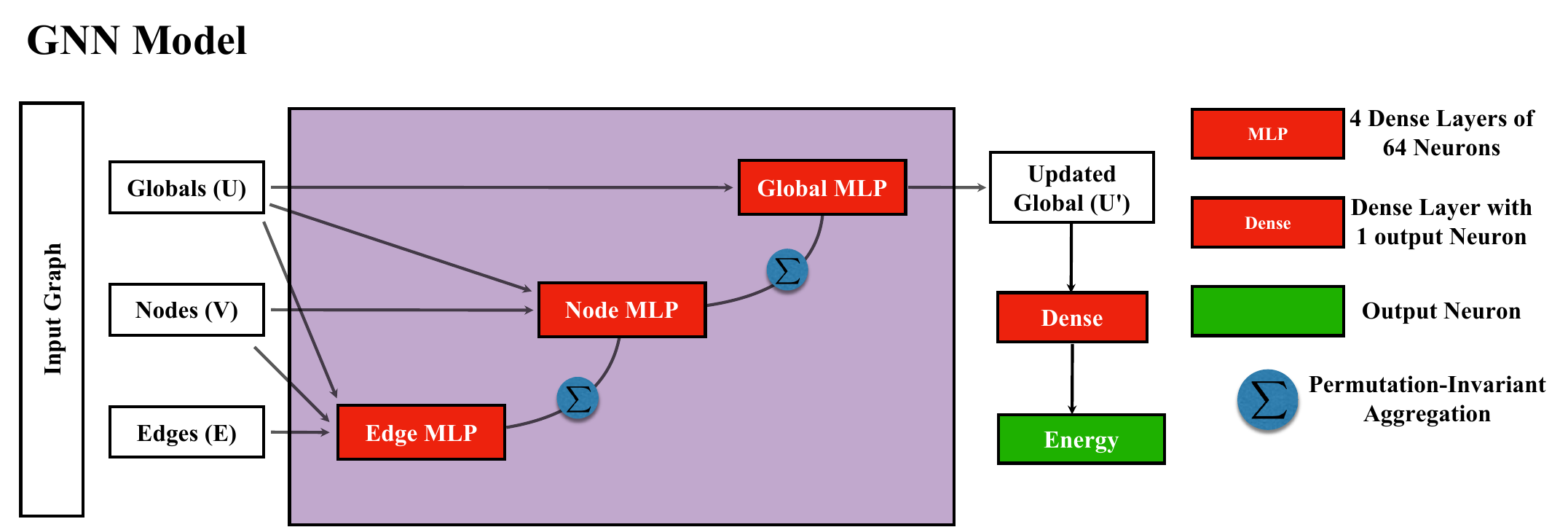}
        \end{center}
        \caption{A schematic of the GNN model.}
        \label{fig:GNN_block}
    \end{figure*}

The functions $f_\mathrm{edge}$, $f_\mathrm{node}$, and $f_\mathrm{global}$ are approximated by \textit{MLPs} as described in Sec. \ref{sec:DeepSets}. Figure \ref{fig:GNN_block} shows a schematic of the GNN model used in this study. The schematic is very similar to Figure \ref{fig:DeepSets_Block}, but includes an additional Edge MLP.
\\

The \textit{Edge MLP} approximates the $f_\mathrm{edge}$ functions and uses the edge, node, and global features as inputs as shown in Eq. \ref{eq:msg_gnn_edge}. The output of the Edge MLP is aggregated across all edges connected to a node and used along with that node's features and global features as an input to the \textit{Node MLP} which approximates $f_\mathrm{node}$ as shown in Eq. \ref{eq:msg_gnn_nodes}. The output of the Node MLP is then aggregated across all nodes and used along with the global features as an input to the \textit{Global MLP} approximating $f_\mathrm{global}$ given by Eq. \ref{eq:gnn_global_nodes}. The updated edges, nodes, and global features can be used to define the updated output graph. This updated graph can then be used to repeat the message-passing scheme several times by stacking such that the latent graph representation is updated based on information propagated through the whole graph. However, to match the size of the DeepSets and GNN models as closely as possible we only use one of the GNN blocks. As in the case of DeepSets model, the updated global node features $u'$ with a dense layer are used to predict the Energy. The GNN model is trained with the same initial learning rate optimizer, batch size, and number of epochs as the DeepSets model described in Section \ref{sec:DeepSets}

\section{Simulations}
\label{sec:sim}
We employed the \textsc{DD4HEP} framework~\cite{Frank:2014zya} to run \textsc{Geant}~\cite{GEANT4:2002zbu} simulations of a calorimeter system that incorporates both an ECAL and an HCAL. The detector design and geometry are similar to the forward calorimeter system intended for the future ePIC detector at the EIC~\cite{Bock:2022lwp}. Both the ECAL and HCAL feature a non-projective geometry, with tower elements aligned parallel to the beam axis.

The ECAL employs tungsten-powder/scintillating fiber technology~\cite{Tsai:2015bna,sPHENIX:2017lqb}, following the design outlined in Ref.\cite{ATHENA:2022hxb}. Its towers have an area of 10$\times$10 cm$^{2}$, it is 17 cm deep (22 $X_{0}$) and lack longitudinal segmentation. Conversely, the HCAL adopts a design similar to that of CALICE SiPM-on-tile design~\cite{CALICE:2022uwn}, following a geometry proposed in Ref.~\cite{Bock:2022lwp}. It features a 4 mm thick scintillator layer sandwiched between 16 mm absorbers. The system comprises 64 layers of absorbers: the initial four layers are made of tungsten, succeeded by 60 layers of steel, for a total of 6.2$\lambda$. 
The transverse area of the scintillator measures 5$\times$5 cm$^{2}$. 

The ePIC HCAL will employ a method that combines various SiPM-on-tile signals to delineate longitudinal segments within the calorimeter. This will be accomplished by summing SiPM-on-tile signals to establish a nominal granularity of 7 equidistant segments, as described in Ref.~\cite{Bock:2022lwp}. In our study, we examine additional possible configurations by incorporating hit signals to define longitudinal segments. To achieve this, we group the hits from the previously mentioned configuration, which represents the maximum achievable granularity, with each layer being read out independently, resulting in a total of 64 layers

While the described geometry does not capture all details of the actual ePIC mechanical design, it serves as a reasonable approximation for the purposes of this study.

No additional material in front of the calorimeter system is included in the simulation. Signals are digitized with 13-bit ADC, and the simulation does not incorporate electronic noise. The simulation framework, which includes the digitization process, has been validated by replicating CALICE data~\cite{CALICE:2012eac}, as elaborated in previous research~\cite{hcalInsert}.

\subsection{Datasets}
\label{sec:dataset}
 Positive pions ($\pi^{+}$) were generated with a polar angle, $\theta$, in the range $10^\circ < \theta < 30^\circ $. Because the detector is symmetric about azimuthal angle, $\phi$, the particles are generated with an azimuthal angle range of $0^\circ < \phi < 360^\circ$. We generated $\pi^{+}$ particles in uniform log10 space in the range of $1 < E_\text{Truth} < 140$ GeV. A total of 1.5 million single-$\pi^{+}$ showers were generated. 
 
\begin{figure*}[t!]
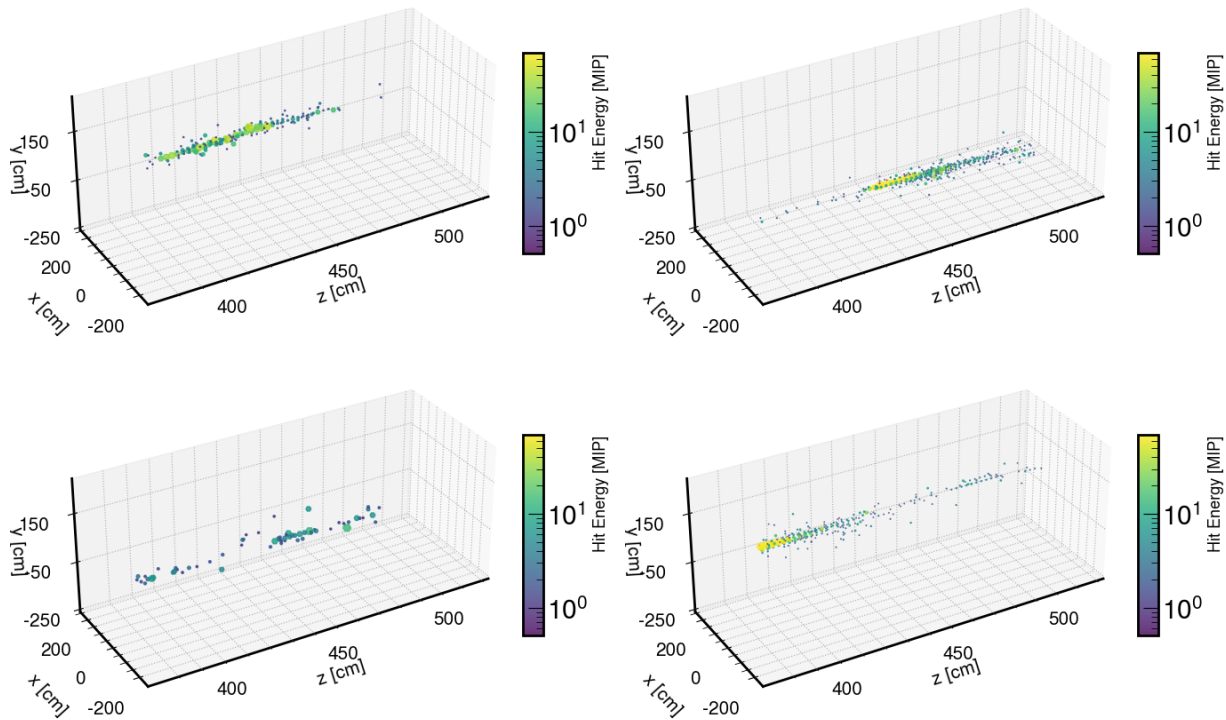

    \centering
        \includegraphics[width=.45\textwidth]{Figures/shower_1.pdf}
        \includegraphics[width=.45\textwidth]{Figures/shower_6.pdf}\\
        \includegraphics[width=.45\textwidth]{Figures/shower_4.pdf}
        \includegraphics[width=.45\textwidth]{Figures/shower_3.pdf}
    \caption{Examples of 4 typical reconstructed 3D shower shapes in HCAL for $\pi^{+}$. The color code represents deposited energy in terms of $E_\text{MIP}$. }
    \label{3D_showers}
\end{figure*}

 Figure~\ref{3D_showers} shows some example shower shapes for $\pi^+$ with random energy, $\theta$, and $\phi$ . As expected, the showers contain two main components: a narrow core with large energy deposits from electromagnetic sub-showers and a halo comprised of low-energy cells from hadronic sub-showers.

The \textsc{Geant} data is stored in files containing two datasets, \textit{clusters} and \textit{cells}. The cluster dataset contains the $E_\text{Truth}$ of the incident pion, as well as the number of hits in the calorimeter. The cell data is  comprised of a constant number of 200 cells per event. The cell data provides energy, position ($X$, $Y$, and $Z$), and time of each hit, represented as point clouds. For studies including ECAL information, an index $k$ is included that indicates whether the cell is part of the ECAL ($k=0$) or part of the HCAL ($k=1$).  In addition to node feature, cluster sum given by the Eq.~\ref{equation_strawman} is provided as the global feature.

Following previous CALICE studies~\cite{CALICE:2012eac,CALICE:2022uwn}, the selections listed below are applied to the data:
\begin{itemize}
    \item Cell energy for each hit must be greater than half of the Minimium Ionizing Particle (MIP) energy for respective calorimeters ($E_\text{cell}> 0.5 \times E_{\text{MIP}}$ ).
    \item Hit time less than 150 ns.
\end{itemize}
$E_\text{MIP}$ is 260 MeV and 130 MeV for the ECAL and HCAL, respectively. $E_\text{MIP}$ is estimated using the most probable value from a 20 GeV $\mu^-$ simulation.
 
The baseline approach, referred to as ``Strawman'', represents the simplest form of reconstruction, which involves summing up the hit energy and accounting for the sampling fraction. Using the Strawman approach, the reconstructed energy can be expressed as follows:
\begin{equation}
    E_\text{Reco} = \frac{\sum^\text{cell} E_i} {SF_\text{hcal}}  + \frac{\sum^\text{cell} E_i} {SF_\text{ecal}},  
    \label{equation_strawman}
\end{equation}
where $SF_\text{hcal}= 2.2\%$, $SF_\text{ecal} = 3.0\%$ are the sampling fraction for the HCAL and ECAL, respectively. Sampling fractions are computed using an electron at fixed energy (40 GeV) as given by Eq.~\ref{equation_SF}:
\begin{equation}
    \textrm{Sampling Fraction}=\left ( \frac{\sum^\text{cell}{ E_i}}  {{E_\text{Truth}}} \right) _{\textrm{at 40 GeV electron}}
    \label{equation_SF}.
\end{equation}

The reported energy scale is the mean obtained through a Gaussian fit to the $E_\text{Reco}/E_\text{Truth}$ distribution in the range of $\pm 3$ standard deviations around the mean value. Resolution is reported as the ratio of sigma to mean from the same Gaussian fit.

\section{Results}
\label{sec:results}
We present results with models described in Sec.~\ref{sec:ML} using progressively more detailed information. We begin with a model trained solely on the sum of energy within the longitudinal layers of the HCAL. Subsequently, we train DeepSets models on granular, cell-level information, using point clouds that incorporate each cell's energy ($E$) and $X$, $Y$, and $Z$ coordinates. Finally, a GNN is trained on comprehensive cell-level information, including edges that represent each cell's nearest neighbors. This approach allows us to evaluate model performance, ranging from a single scalar value (the energy sum) to a GNN that directly encodes the geometry of the energy shower.  To gauge the extent of improvement achieved through ML-based results, we compare them with the Strawman, which represents the most basic form of reconstruction.

 First, the impact of longitudinal segmentation on energy resolution is presented. For this study, we divide calorimeter into distinct $Z$-sections and assessed the resulting resolution. For instance, in case of $n^{th}$ $Z$-section, the calorimeter was longitudinally divided into $n$ equidistant $Z$-sections. The cell energy is integrated within each $Z$-section for each unique combination of $X$ and $Y$. The new $Z$ position was determined as the center of these $Z$-sections. The regrouped hit information was subsequently utilized as input for the DeepSets model. 
 
\begin{figure*}[!h]
    \centering
	\includegraphics[width=0.49\textwidth]{Figures/z_section.pdf}
	\includegraphics[width=0.5\textwidth]{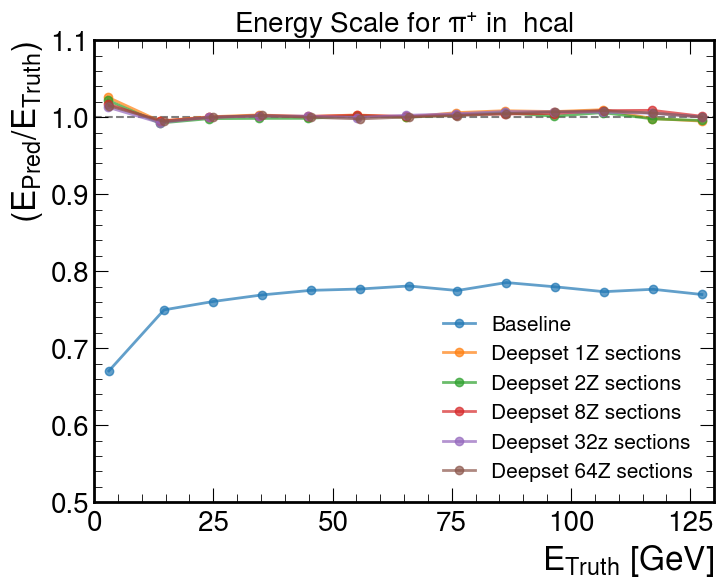}
	\caption{Energy resolution (left) and energy scale (right) of calorimeter with different number of $Z$-sections along the longitudinal direction. The bottom panel of resolution plot shows the square root of difference in squares of resolution of 1 $Z$-section and the given $Z$-sections.}
    \label{performance_zSeg}
	\end{figure*}

 Figure~\ref{performance_zSeg} (left) shows the resulting energy resolution and energy scale (right) for various $Z$-sections. The poor performance of the baseline reconstruction (represented by magenta open circles) indicates that relying solely on simple energy reconstruction is inadequate for addressing energy regression with the ECAL and HCAL when they have different $e/h$ ratios. The resolution improves with an increasing number of $Z$-sections using the DeepSets model.  The bottom panel on the resolution plot shows the square root of the difference in squares of resolution of one $Z$-sections and the given $Z$-section as a measure of improvement of resolution relative to one $Z$-section. Improvement in resolution for $Z$-sections is dependent on energy; lower energies exhibit the greatest improvement.

An enhancement of approximately 20\% in resolution is attainable when using eight $Z$-sections as opposed to just one $Z$-section. The Ref.~\cite{Bock:2022lwp} describes the default HCAL configuration with seven $Z$-sections. The DeepSets model demonstrates superior performance, even with the minimum number of $Z$-sections, compared to the most basic reconstruction method.

\begin{figure*}[!h]
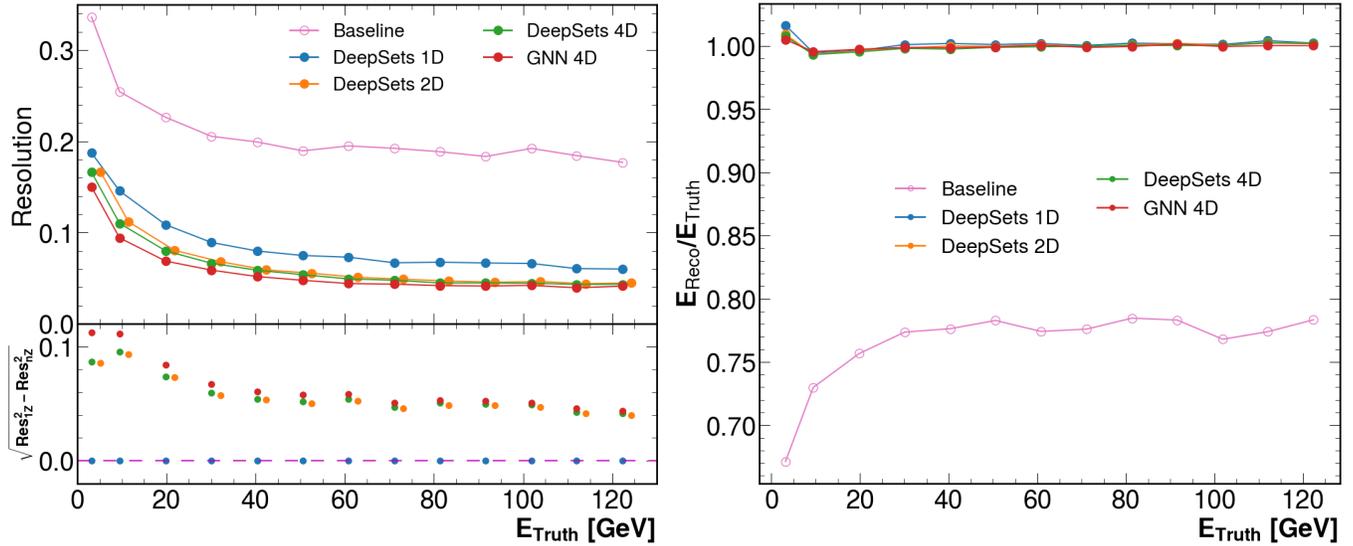

    \centering
	\includegraphics[width=0.490\textwidth]{Figures/res_graphnet_deepset.pdf}
	\includegraphics[width=0.50\textwidth]{Figures/scale_graphnet_deepset.pdf}
	\caption{Comparison of regression model performance with different input features. Energy resolution (left) energy scale (right). A 2 GeV shift has been applied to the energy values along the $X$-axis for the 2D data.}
    \label{performance_input_dimension}
	\end{figure*}

The next set of results investigates the regression performance with different levels of input features.
Figure~\ref{performance_input_dimension} shows the energy regression performance of models trained on different input features. As described in Sec.~\ref{sec:DeepSets}, the 1D DeepSets model is trained only on the hit energy. The 2D DeepSets is trained on hit energy and $Z$, while the 4D DeepSets model is trained on hit energy, $X$, $Y$, and $Z$ information from all cells.  When we incorporate longitudinal information in addition to hit energy into the DeepSets models, we observe about a 30\% improvement in the energy resolution (1D model vs. 2D model). Conversely, the inclusion of transverse information does not appear to influence the performance significantly (2D vs. 4D Model). 

We found the GNN model improves the energy resolution by 10\%  compared to the 4D DeepSets model. This indicates that the edges encode valuable information that improves regression performance beyond a DeepSets model trained on the full point cloud representation of the shower.  Note that the DeepSets and GNN models are both universal for point clouds---edges are not formally required to achieve the optimal performance, but for our non-asymptotic setup, they seem to be useful (indicating effective inductive bias). The bottom panel of resolution shows the improvement in resolution relative to 1D DeepSets model.

    \section{Conclusions and Outlook}
    \label{sec:conclusions}
We investigated the influence of longitudinal and transverse calorimeter segmentation on energy regression utilizing DeepSets and GNNs with a detector design similar to the one intended for the upcoming EIC. This allows us to achieve optimal reconstruction performance for a given geometry, which is a prerequisite for determining the optimal detector configuration.

We show the DeepSets model with minimal input outperforms  the performance of our baseline algorithm. We also compare the calorimeter performance with different input features and different models. We found 2D input ($E$, $Z$) improves resolution by about ~30\% compared to 1D input features (only $E$). Interestingly, no significant improvement from 2D ($E$, $Z$) to 4D ($E$, $Z$, $X$, $Y$) input features is observed, indicating that the longitudinal information of the hadronic shower is more useful for precise energy regression than transverse information. We also studied the impact of longitudinal information on its own in calorimeter resolution. We found that increasing the number of longitudinal sections improves the calorimeter performance significantly, especially at energies below 30 GeV. With 64 $Z$-sections (maximum limit for our setup), one can achieve $\approx$ 50\% improvement in resolution relative to one $Z$-section. We also presented a comparison between two models: DeepSets and GNNs. The GNN model improves resolution by about 10\% compared to DeepSets with similar input. 

This study demonstrates the effectiveness and versatility of ML-based regression using point-cloud representation for high-granularity calorimeters. It reveals that these systems can achieve performance improvements of up to a twofold increase in resolution while achieving excellent energy scale and linearity. We show that this applies to realistic calorimeter setups. Even if such a system lacks longitudinal segmentation in the ECAL, we observe significant enhancements in the performance of hadronic showers across the entire phase space relevant to the EIC when fine longitudinal segmentation is employed in the HCAL.

 The quantitative analysis presented here regarding the impact of longitudinal segmentation, whose performance gains depend on energy, serves as motivation and guidance for optimizing the granularity of the EIC forward calorimeter based on pseudorapidity. The implementation of such an approach is feasible thanks to the flexibility of SiPM-on-tile technology, which can be employed to define tower sections of varying sizes.

This research is part of a larger movement to automate detector design with machine learning.  Recent works have considered a number of different instruments with a variety of approaches~\cite{Cisbani:2019xta,Fanelli:2022rdm,Strong:2023oew,MODE:2022znx,Fanelli:2022kro,hepmllivingreview}.  The studies presented here could be combined with a point cloud generative model for an end-to-end optimization~\cite{Kansal:2021cqp,Buhmann:2023pmh,Kach:2022qnf,Verheyen:2022tov,mikuni:point_clouds,Leigh:2023toe,Acosta:2023zik}.  Such an approach could prove beneficial for several high-granularity designs at the EIC that are currently under optimization, including the forward calorimeter system, the barrel electromagnetic calorimeter, and the zero-degree calorimeters.  A full hadronic final state reconstruction at the EIC will likely integrate calorimeteric and tracking information in an energy-flow algorithm---future studies could also include this into a global optimization.

\section*{Code Availability}
 The code for the data processing, training models, and plotting results can be found here: \url{https://github.com/eiccodesign/regressiononly}. The data used for these studies can be found here: \url{https://zenodo.org/record/8384822}

    \section*{Acknowledgments}
We acknowledge support from DOE grant award number DE-SC0022355. This research used resources from the LLNL institutional Computing Grand Challenge program and the National Energy Research Scientific Computing Center, a DOE Office of Science User Facility supported by the Office of Science of the U.S. Department of Energy under Contract No. DE-AC02-05CH11231 using NERSC award HEP-ERCAP0021099. M.A acknowledges support through DOE Contract No. DE-AC05-06OR23177 under which Jefferson Science Associates, LLC operates the Thomas Jefferson National Accelerator Facility. This work was performed under the auspices of the U.S. Department of Energy by Lawrence Livermore National Laboratory under Contract No. DE-AC52-07NA27344 and by Lawrence Berkeley National Laboratory under Contract No. DE-AC02-05CH11231.

    %BN is supported by the U.S. Department of Energy (DOE), Office of Science under contract DE-AC02-05CH11231. 

    \bibliographystyle{apsrev4-1}
    \bibliographystyle{unsrt} % or try abbrvnat or unsrtnat
    
    \bibliography{biblio}

    \end{document}